\documentclass[twocolumn,showpacs,preprintnumbers,amsmath,amssymb]{revtex4}

\usepackage{graphicx}
\usepackage{dcolumn}
\usepackage{bm}

\begin{document}

\title{Parametric investigation of nonlinear fluctuations in a dc glow discharge plasma}%
\author{Md. Nurujjaman}
\email{md.nurujjaman@saha.ac.in}
\author{Ramesh Narayanan}

\author{A.N. Sekar Iyengar}
\email{ansekar.iyengar@saha.ac.in}

\affiliation{Plasma Physics Division,
Saha Institute of Nuclear Physics,
1/AF, Bidhannagar, 
Kolkata -700064, India.}
\begin{abstract}Glow discharge plasmas exhibit various types of self excited oscillations for different initial conditions like discharge voltages and filling pressures. The behavior of such oscillations associated with the anode glow have been investigated using nonlinear techniques like correlation dimension, largest Lyapunov exponent etc. It is seen that these oscillations go to an ordered state from a chaotic state with increase in input energy i.e. with discharge voltages implying occurrence of inverse bifurcations. These results are different from the other observations wherein the fluctuations have been observed to go from ordered to chaotic state.
\end{abstract}

\maketitle

\section*{}
\textbf{Chaos and different route to chaos have been studied in many glow discharge plasma system. In the earlier experiments which are mainly performed in the parallel plate electrode systems, it is observed that the system becomes chaotic with increasing input energy. But in cylindrical system, we have observed an opposite phenomena i.e.  chaotic to stable state transitions with increase in input energy, where the energy is given the form of the discharge voltages. This system has extensive use in the dusty plasma experiments. We provide a qualitative and quantitative description, based on ideas and methods from nonlinear time series analysis, of transitions in these dynamics. We also provide a plausible model for this phenomenon. We hope this work can help improve the fundamental understanding of such inverse phenomenon.}

\section{Introduction}
Plasma is a typical complex medium exhibiting a wide variety of nonlinear phenomena such as self oscillations, chaos, intermittency etc~\cite{prl:Ding,prl:Ding1,pop:Hassouba,pramana:jaman}. These systems exhibit chaotic or regular behavior depending on the controlling parameters like discharge current, voltage, pressure etc, and the transition from one state to another takes place for a change in any one of the parameters~\cite{prl:Cheung,prl:Qin,pop:Hassouba,EurPhysJD:Atipo,prl:braun}. In this paper we report on the observations of the chaotic and the regular behaviors in the floating potential fluctuations at different discharge voltages in a concentric cylindrical electrode system with a hollow cathode. In almost all the earlier experiments, the system goes from order state to chaotic state, whereas in ours the reverse is seen, thus indicating signature of inverse bifurcation for the first time in a glow discharge plasma. The fluctuations are observed when a bright glow is formed around the anode. Nonlinear fluctuation analysis has been carried by estimating the correlation dimension and Lyapunov exponent in addition to power spectral analysis. The correlation dimension and Lyapunov exponents were estimated using the Grassberger-Procaccia algorithm~\cite{physrevlett:grassberger}, the Rosenstein's technique algorithm~\cite{physicaD:rosenstein}. The latter was chosen for the Lyapunov exponent since it is efficient for small data lengths~\cite{physicaD:rosenstein,PhysicaD:wolf}.

\section{Experimental setup }
\label{subsection:Experimental setup }
\begin{center}
\begin{figure}[ht]
\includegraphics[width=8.5 cm]{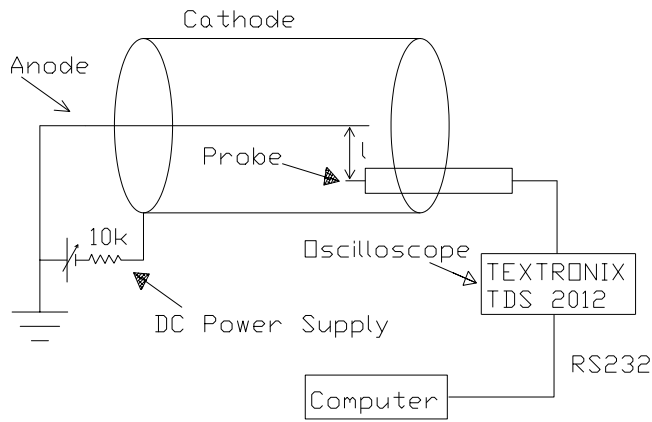}
\caption{Schematic diagram of the cylindrical electrode system of the glow discharge plasma. The probe is placed at a distance $l \approx12.5$ mm from the anode. }
\label{fig1:setup}
\end{figure}
\end{center}

The experiments were performed in a hollow cathode dc glow discharge plasma. The schematic diagram is shown in Fig~\ref{fig1:setup}. A hollow S.S. tube of diameter ($\phi$) $\approx$ 45 mm was the cathode and a central rod of $\phi\approx$ 1.6 mm was the anode. The whole assembly was mounted inside a vacuum chamber and was pumped down to about  0.001 mbar using a rotary pump. The chamber was filled with argon gas up to a desired  value of neutral pressure, by a needle valve, and then a discharge was struck by a dc voltage (which we called as discharge voltage or DV), which could be varied in the range of 0$-$1000 V. The electrostatic floating potential fluctuations were measured using a Langmuir probe of $\phi$ = 0.5 mm  and 2 mm long, approximately at the mid position of the anode and the cathode. A black and white CCD camera has been used to  measure the size of the anode glow. The measured plasma density and the electron temperature were about $10^7cm^{-3}$ and 3$-$4 eV respectively. The corresponding electron plasma frequency ($f_{pe}$), and the ion plasma frequency ($f_{pi}$) were about 28 MHz, and 105 kHz respectively. 

\section{Results }
\label{subsection:results}
\subsection{Model for anode glow}

\begin{center}
\begin{figure}[ht]
\includegraphics[width=8.5 cm]{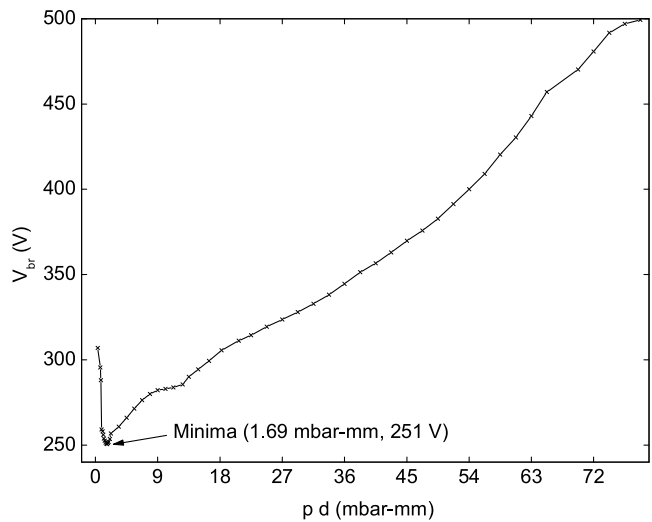}
\caption{ $V_{br}$ vs $pd$ (Paschen curve) for our experimental system. The minimum occurs at (1.69 mbar-mm, 251 V).}
\label{fig:Paschen}
\end{figure} 
\end{center}

By varying the neutral pressure, we observed the breakdown of the gas at different voltages as shown in Fig~\ref{fig:Paschen}. The breakdown voltage ($V_{br}$) initially decreases with increase in $pd$, (where p and d are the filling pressure and radius of the cathode respectively)  and then begins to increase with $pd$ after going through a minimum value resembling  a typical paschen curve~\cite{book:von}. 
But the characteristics of the discharges are very different at low and high pressures. At higher pressures especially for $pd >20$ mbar-mm an anode glow was observed as shown in fig 3.
 
\begin{center}
\begin{figure}[ht]
\includegraphics[width=8.5 cm]{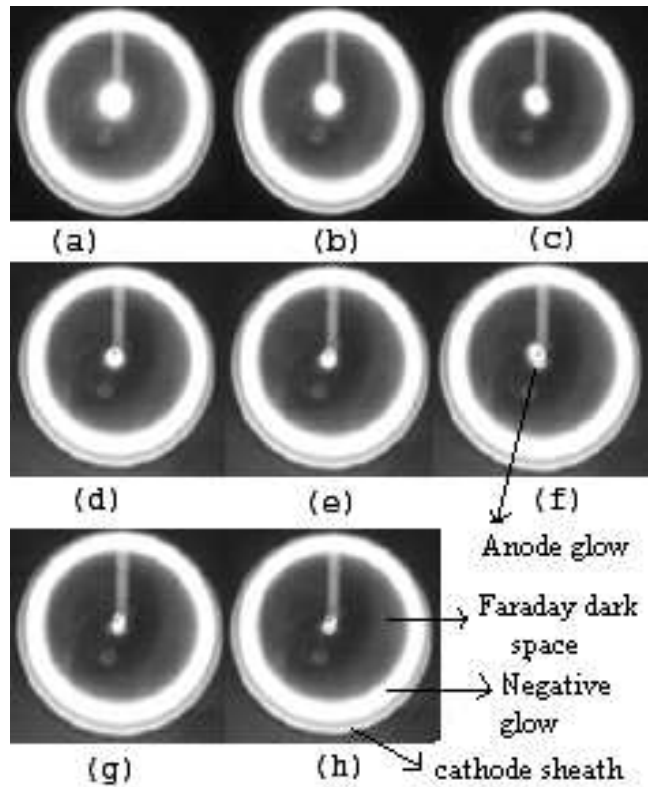}
\caption{Evolution of glow size of the anode glow with increasing DV (a) - (h).}
\label{fig:glow}
\end{figure} 
\end{center}

Figs ~\ref{fig:glow}(a) shows that the glow with largest size, appears when the discharge is struck at a typical pressure of 0.95 mbar and its size decreases with increase in the DV until it finally disappears [Figs~\ref{fig:glow}(a)$-$\ref{fig:glow}(h)]. From the CCD image analysis the annular radius of the glow around the anode, was estimated to be $\approx 1.3$ mm at the beginning of the discharge [Fig~\ref{fig:glow}(a)] and reduced to $\approx 0.32$ mm [Fig~\ref{fig:glow}(f)]. We estimated the thickness ($\delta$) using the relation~\cite{JphysD:song} 
\begin{equation}
\label{eqn:thickness}
\delta\approx3.7\times10^{-6}\frac{kT}{\sigma_{i}P},
\end{equation}
 where k, T, $\sigma_i$, and P are Boltzmann constant, room temperature in Kelvin scale, ionization cross section and presure in mbar respectively. $\sigma_i$ depends upon the energy of the electron and at 15.76 eV, $\sigma_i$ is $\approx2\times10^{-18}cm^2$~\cite{jchemphys:rapp}. The estimated thickness of the anode glow for $P=0.95$ mbar; $T=300$ K, is $\approx0.81$ mm which is  within the range of the thickness estimated from the image shown in Fig~\ref{fig:glow}. 

The anode glow appears when the discharge current is too low to sustain the discharge~\cite{ppcf:sanduloviciu,JphysD:song}. This glow supplies energy, in the form of positive potential gradient, to the electrons so that additional ionization takes place and hence discharge current increases~\cite{ppcf:sanduloviciu,JphysD:song}. In the present experiment, at the initial stages of discharges the electrons do not have sufficient energy near the anode to reach it, and hence the anode glow occurs. But, when the electrons have sufficient energy to reach the anode the glow disappears.

\begin{center}
\begin{figure}[ht]
\includegraphics[width=8.5 cm]{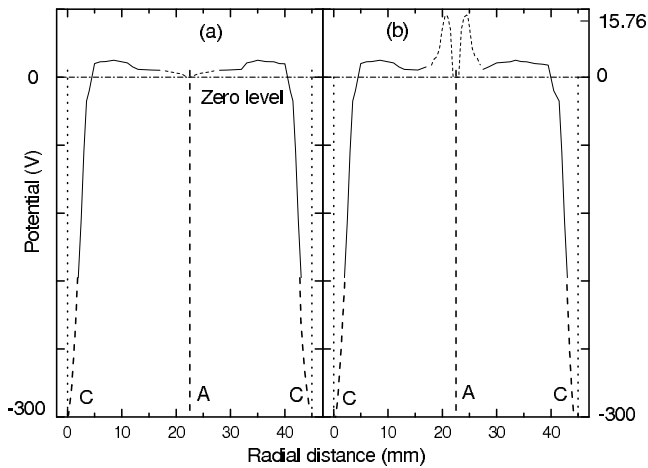}
\caption{Spacial variation of the plasma floating potential: (a) without anode glow; (b) with anode glow. Vertical dotted  and dashed lines are the cathode positions (C) and anode position (A) respectively. Horizontal dash dot line is the zero level. $--$ in the curve near the anode and the cathode are the extrapolation of the potential.}
\label{fig:pot}
\end{figure} 
\end{center}

Fig~\ref{fig:pot}(a) and~\ref{fig:pot}(b) show the tentative model of floating potential profile for the present experimental arrangement. It shows that the floating potential increases sharply in the sheath region near the cathode and then decreases slowly up to the anode. As the measurement near the anode and the cathode was not possible, the extrapolated potential profile is shown by dash lines. In the the presence of the anode glow, the potential near the anode get modified. Experimentally it is observed that the anode glow is separated from rest of the plasma by a double layer~\cite{ppcf:sanduloviciu,JphysD:song,JphysD:opresu,pla:sanduloviciu} and the height of the potential of the core of this glow is of the order of the ionization potential~\cite{JphysD:opresu}, which for argon is about 15.76 eV. As the anode is grounded, the modified potential profile in this case, over the profile shown in Fig~\ref{fig:pot}(a), will be as shown in Fig~\ref{fig:pot}(b). As the width of the anode glow decreases with increase in the DVs [Fig~\ref{fig:glow}] the hump shown in Fig~\ref{fig:pot}(b) will also shrink accordingly. This shrinkage happens because as the electron energy increases with the DVs, the necessity of the additional ionization to maintain the discharge decreases.

\subsection{Analysis of the floating potential fluctuations}
An interesting feature associated with the anode glow was the different types of oscillations in the floating potential at different pressures.

\begin{figure}[ht]
\includegraphics[width=8.5 cm]{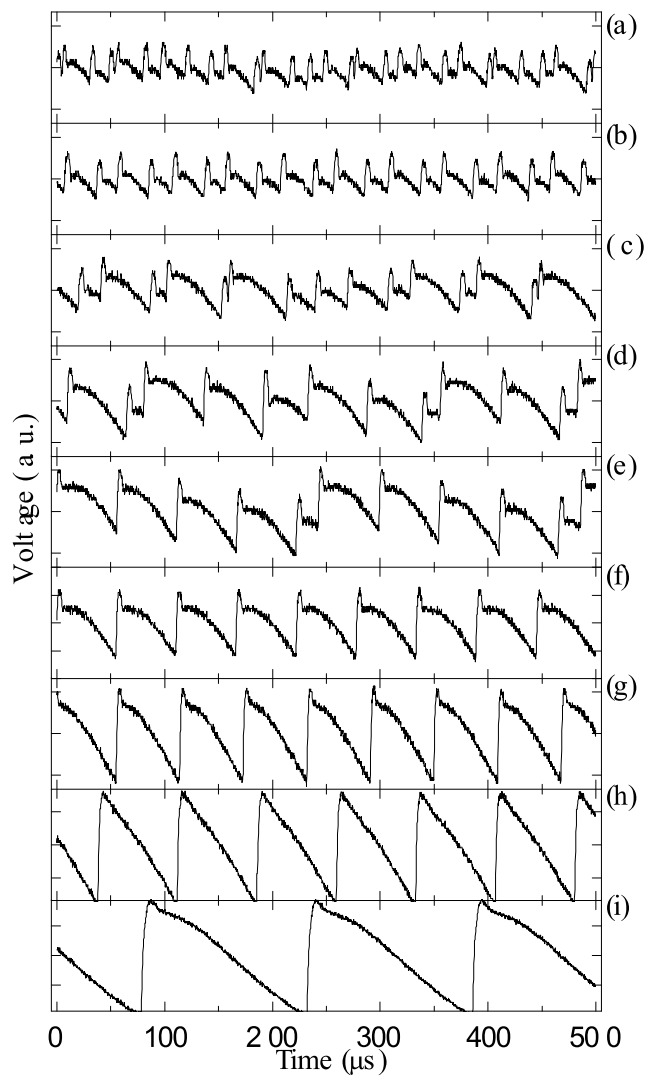}
\caption{Sequential change in the raw signal (normalized) at 0.89 mbar for different voltages: (a) 288 V; (b) 291 V; (c) 295 V; (d) 301 V; (e) 304 V; (f) 307 V; (g) 327 V; (h) 385 V; (i) 466 V. All y-axes range form -1 to 1.}
\label{fig:raw0.89mb}
\end{figure} 
\begin{figure}[ht]
\includegraphics[width=8.5 cm]{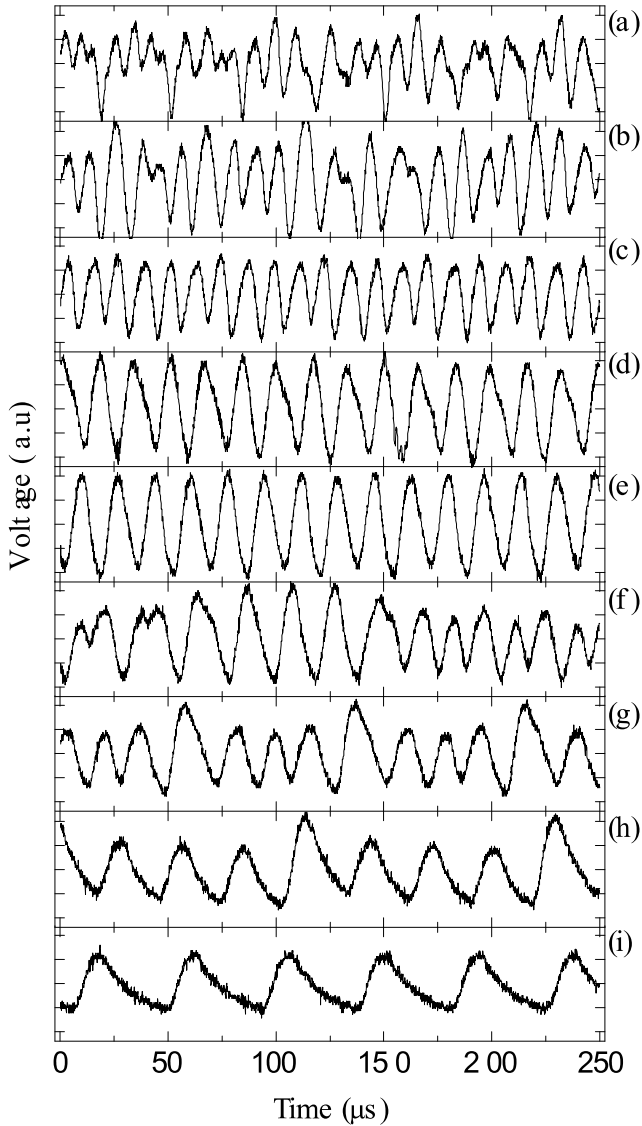}
\caption{Sequential change in the raw signal (normalized) at 0.95 mbar for different voltages:(a) 283 V; (b) 284 V; (c) 286 V; (d) 288 V; (e) 289 V; (f) 290 V; (g) 291 V; (h) 292 V; (i) 293 V. All y-axes range form -1 to 1.}
\label{fig:raw0.95mb}
\end{figure} 

\begin{figure}[ht]
\includegraphics[width=8.5 cm]{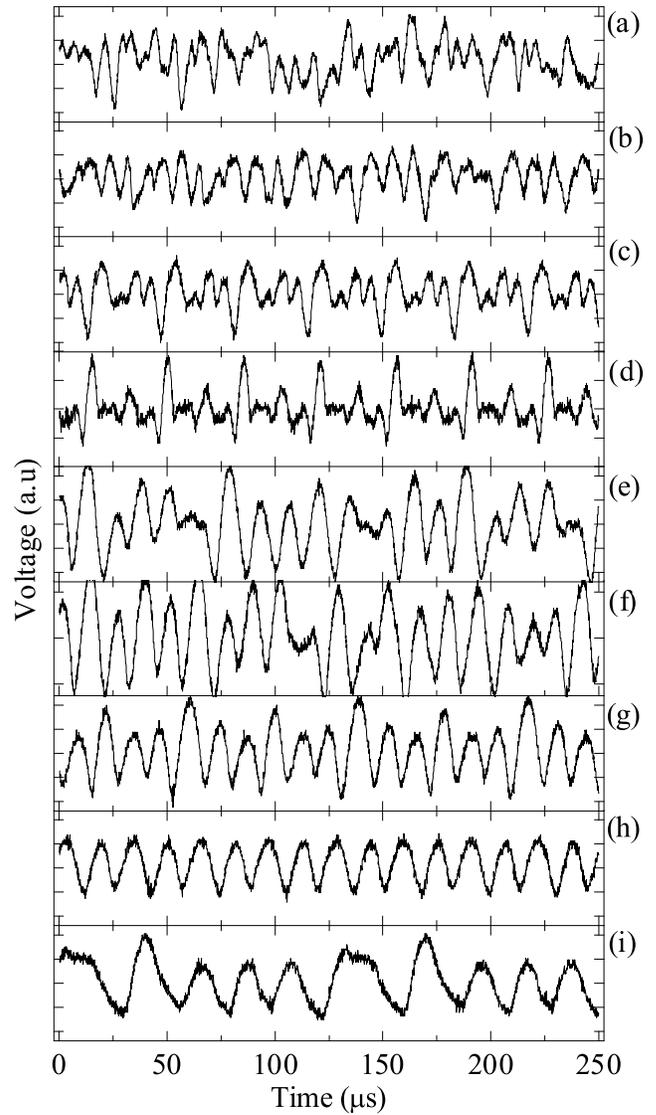}
\caption{Sequential change in the raw signal (normalized) at 1mbar for different voltages: (a) 293 V; (b) 296 V; (c) 298 V; (d) 299 V; (e) 300 V; (f) 305 V; (g) 308 V; (h) 310 V; (i) 312 V. All y-axes range form -1 to 1.}
\label{fig:raw1mb}
\end{figure} 

The behavior of the floating potential fluctuations for pressures less than the Paschen  minimum, was different from that at higher pressures. At low pressures (LHS of Paschen minimum) the fluctuations exhibited Self Organized Criticality (SOC) like behavior~\cite{Physleta:jaman}, but at higher pressures (RHS of the Paschen minimum), we observed fluctuations in the floating potential with the formation of the anode glow (fire ball, similar to Ref~\cite{ppcf:sanduloviciu,JphysD:song}) which also disappears with the disappearance of the glow [Fig~\ref{fig:glow}].

We carried out a detailed analysis of the fluctuations for three typical pressures which are presented here. At about 0.89 mbar ($pd\approx 20.02$ mbar-mm), the discharge was initiated at $\approx 288$ V and an anode glow was observed similar to Fig~\ref{fig:glow}(a). Simultaneously irregular relaxation type of oscillations in the floating potential were observed [Fig~\ref{fig:raw0.89mb}(a)]. Increasing the DVs, led to an increase both in the amplitude and the time period of the oscillations [Figs~\ref{fig:raw0.89mb} (b)$-$\ref{fig:raw0.89mb}(i)]. The regularity of the oscillation also increases with the DVs. Around 509 V both the anode glow and the fluctuations disappeared simultaneously. At 0.95 mbar ($pd\approx 21.37$ mbar-mm) the fluctuations are observed to be more random [Fig~\ref{fig:raw0.95mb}(a)$-$(i)] than at 0.89 mbar. With increasing DV, the final form of the fluctuations before their disappearance was the relaxation type of oscillation as shown in Figs~\ref{fig:raw0.95mb}(h)$-$\ref{fig:raw0.95mb}(i). Increasing the pressure causes more randomness in the signals as seen in Fig~\ref{fig:raw1mb}(a)$-$(i) at 1.0 mbar ($pd\approx 22.5$ mbar-mm). We observed the relaxation type of oscillations  at DV $\approx313$ V just before the glow an the fluctuation disappear. 

\begin{figure}[ht]
\includegraphics[width=8.5 cm]{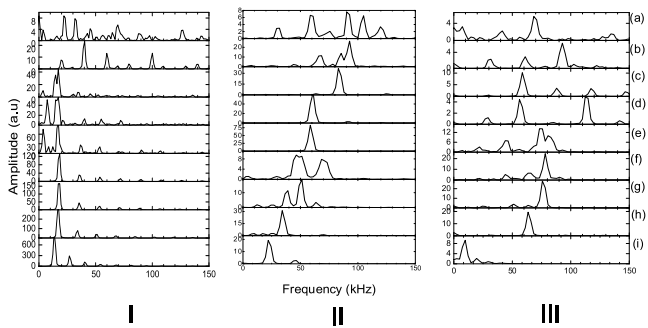}
\caption{Power spectrum of floating potential fluctuations at initial discharge voltages at  filling pressures: (a) 0.89 mbar, (b) 0.95 mbar and (c)1 mbar. }
\label{fig:powspecs}
\end{figure}

\begin{center}
\begin{figure}[ht]
\includegraphics[width=8.5 cm]{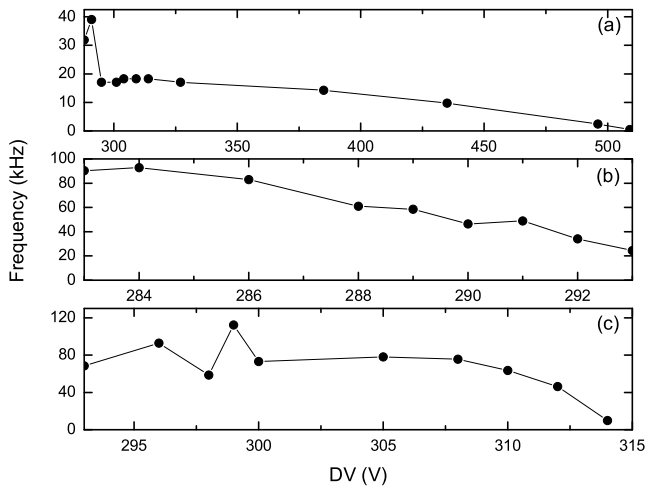}
\caption{Dominant frequency vs DV at the three experimental pressure: (a) 0.89 mbar; (b) 0.95 mbar; (c) 1.0 mbar.}
\label{fig:domifreq}
\end{figure} 
\end{center}

Figs~\ref{fig:powspecs}\{I\}, \{II\} and \{III\} show the power spectrum calculated for the signals at three different experimental pressures shown in Figs~\ref{fig:raw0.89mb},~\ref{fig:raw0.95mb} and ~\ref{fig:raw1mb} respectively. A broad range of frequencies ($<$ ion plasma frequency) is observed at the initial stage of the discharge, at pressures 0.89 mbar [Fig~\ref{fig:powspecs}\{I\} (a)], 0.95 mbar [Fig~\ref{fig:powspecs}\{II\} (a)] and 1.0 mbar [Fig~\ref{fig:powspecs}\{III\} (a)$-$(b)]. The figures also show that with increasing DV, number of peaks present in the power spectrum, decreases. At higher DV i.e. at the end of the sequence of the power spectra we observed one major peak with its harmonic [Figs~\ref{fig:powspecs}\{I\}(f)$-$(i) and III(i)] or two peaks [Figs~\ref{fig:powspecs}\{II\}(i)] just before the fluctuations cease. The power spectra clearly show that the system goes from an irregular to a more or less regular state with increasing DV. Another interesting feature that has been observed for all the three pressures is that the dominant frequency i.e. the frequency which has maximum power  decreases with increase in the DVs [Figs~\ref{fig:domifreq}(a)$-$(c)].  On the contrary, in many experiments~\cite{pop:lee,pop:Klinger,ppcf:gyergyek}, the dominant frequency is observed to increase with the DV. The basic difference between the present experiment and the experiments performed in Ref~\cite{pop:lee,pop:Klinger,ppcf:gyergyek} was in the electrode configuration and biasing arrangement. In those experiments~\cite{pop:lee,pop:Klinger,ppcf:gyergyek} the system was planar type is which the anode voltage was increased keeping the cathode at ground whereas ours is a cylindrical system in which we increased the negative voltage on the cathode keeping anode at ground. In those systems electrons originating at cathode get lost at the anode. In our case the electrons can undergo multiple reflections and hence increases the ionization length. The fluctuations are probably due to ion acoustic instability driven by electron beam plasma interaction and also from electron and ion trapping in the potential wells. associated with the anode glow [Fig~\ref{fig:pot}].

Ions produced inside the anode glow due to collisions of the accelerated electrons across the hump with the neutrals, makes the glow unstable~\cite{ppcf:sanduloviciu}, which is probably responsible for the relaxation or the random oscillations~\cite{pop:lee,pop:Klinger}. An estimate of the frequency of these instabilities can be obtained from the ion transit time in the plasma~\cite{pop:Klinger}  $\tau(d)=\frac{d}{V_{th,i}}=d/\sqrt\frac{k_bT_i}{m}$, where d is the electrode distance. The estimated ion transit frequency ($\frac{1}{\tau}$) for our experimental system is $\approx19$ kHz which agrees well with the frequency of the relaxation oscillations of the floating potential shown in Figs~\ref{fig:powspecs}\{I\}(f)$-$(i),\{II\}(i) and \{III\}(i). The higher frequency oscillations could be due to ion acoustic instabilities since the conditions are quite conducive to excite these low frequency instabilities. In the next section we have presented nonlinear analysis of the fluctuations. 

The presence of relaxation oscillations have been attributed to the formation of highly nonlinear structures like double layers~\cite{pre:Valentin}. We therefore estimated the correlation dimension ($D_{corr}$) and the +ve Lyapunov exponent ($\lambda_L$) of all the signals. In this experiment, it is observed that the nature of the fluctuations of the potential did not vary for almost the whole day as long as the controlling parameters are kept constant, so the analyzed signals can conveniently be taken to be stationary.  

\begin{figure}[ht]
\includegraphics[width=8.5 cm]{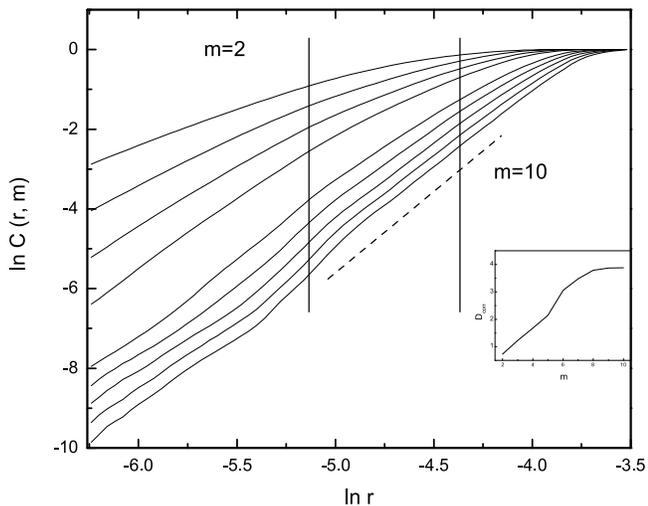}
\caption{ Effect of embedding dimension on correlation sums. The scaling region is shown by two vertical lines. The best fitting is shown by $--$ line.}
\label{fig:corr}
\end{figure}

\begin{figure}[ht]
\includegraphics[width=8.5 cm]{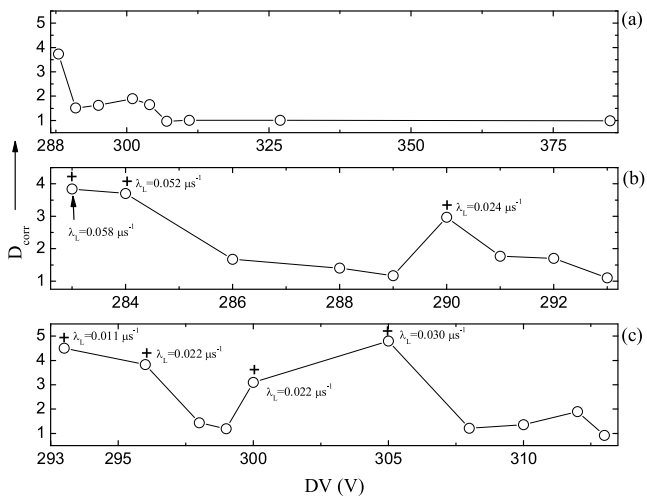}
\caption{$D_{corr}$ vs discharge voltages.  +ve $\lambda_L$ has been shown by +ve sign. at (a) 0.89 mbar; (b) 0.95 mbar; (c) 1 mbar.}
\label{fig:cordisch}
\end{figure}

The correlation dimension ($D_{corr}$) has been calculated using well-known Grassberger-Procaccia techniques~\cite{physrevlett:grassberger,physrevA:grassberger}, where correlation sum (C(r,m)) scales with scale length (r) as $C(r,m)=r^D_{corr}$.  A typical plot of $\ln C(r,m)$ vs $\ln r$ has been shown in  Fig~\ref{fig:corr} for the embedding dimension (m), in the range of 2 to 10, from which we have estimated the correlation dimension ($D_{corr}$) for a typical signal (Fig~\ref{fig:raw0.95mb}(a)) at DV $\approx283$ V,  at  0.95 mbar. The scaling region had to be chosen carefully, since for too small scale lengths the correlation sum is heavily distorted by noise and the higher scale lengths are limited by attractor dimensions. From the above plot the correlation sum exhibits a power law behavior within a certain range of $r$ as shown by the vertical dotted line. $\ln C(r,m)$ vs $\ln r$ plots are almost parallel for higher m (i.e. $m=7-10$) and the corresponding best fit has been shown by $--$ line. $D_{corr}$ vs m is also shown in the inset of Fig~\ref{fig:corr}. It shows that $D_{corr}$ becomes constant at higher m and this constant value of $D_{corr}$ is the correlation dimension of that particular signal and for all of our data $D_{corr}\geq3.8$ to begin with. The $D_{corr}$ at 0.89, 0.95 and 1 mbar for different DVs have been shown by open circle($\circ$) in Figs~\ref{fig:cordisch}(a)$-$\ref{fig:cordisch}(c) respectively. It is observed that there is a decreasing tendency  of $D_{corr}$ except for some intermediate values of DV at higher pressures, where it is seen to increase and then decrease again. Since $D_{corr}$ is a measure of the complexity of the system it is likely that the system complexity increases at those intermediate DVs. In our experiment initially the system is in a complex state as $D_{corr}$ for all the three pressures is high and decreases with increase in DVs [Fig~\ref{fig:cordisch}(a)$-$(c)].  $D_{corr}\approx1$ just before the system reaches stable state finally, is the indicator of the periodic state and these periodic nature are also prominent from the raw data and the power spectrum [Figs~\ref{fig:raw0.89mb},~\ref{fig:raw0.95mb}, ~\ref{fig:raw1mb} and Fig~\ref{fig:powspecs}\{I\}$-$\{III\}]. $D_{corr}$ shows that the system stabilizes itself with increase in DVs.


\begin{figure}[ht]
\includegraphics[width=8.5 cm]{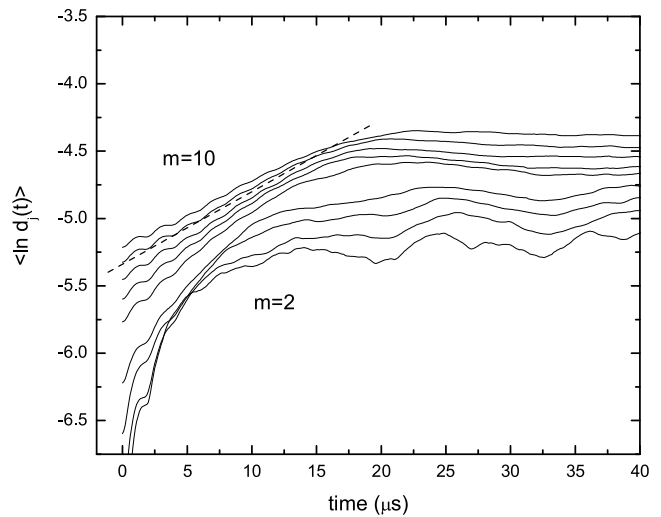}
\caption{ Average $\ln d_j(t)$ for different embedding dimensions. The best region for $\lambda_L$ has been shown by dotted line.}
\label{fig:lyap}
\end{figure}

\begin{table}
\caption{Error in the estimation of the correlation dimension and the +ve Lyapunov exponent}
\label{table:lyap}
\centering
\begin{tabular}{|c| c |c|}
\hline
\hline
data length & $D_{corr}$ &$\lambda_L$ \\
\hline
500 & 9.47 &0.240 \\
1000 & 4.00 &0.056 \\
1500 & 3.85 &0.058 \\
2000 & 3.78 &0.054 \\
2500 & 3.87 &0.058 \\
\hline
\end{tabular}
\end{table}

The presence of a +ve Lyapunov exponent ($\lambda_L$) is the most reliable signature of the chaotic dynamics and it is estimated using the Rosestein algorithm~\cite{physicaD:rosenstein} which is especially suited for small data lengths. $\lambda_L$ can be estimated from the slope of the plot $<\ln d_j(t)>$ vs time ($i\Delta t$) as shown in Fig~\ref{fig:lyap}. The figure shows $<\ln d_j(t)>$ vs $i\Delta t$ for  m=2 to 10. A clear scaling region is seen at a higher m shown by $--$ line. The positive $\lambda_L$ has been identified at 0.95 and 1 mbar for some DVs and they are shown in Fig~\ref{fig:cordisch}(b) and \ref{fig:cordisch}(c) by +ve sign respectively. Figs~\ref{fig:cordisch}(b) and (c) show that $\lambda_L$ becomes positive for  283, 284, and 290 V at 0.95 mbar and for 293, 296, 300 and 305 V at 1 mbar respectively. Though $D_{corr}$ quantifies the complexity present in the system, it does not guarantee the presence of chaos, which is determined by $\lambda_L$ . At  0.89 mbar initially we have a high $D_{corr}$ [Fig~\ref{fig:cordisch}(a)], but the $\lambda_L$ is not positive in this case, implying that the system is not in chaotic state.  At higher pressures we find that in general $\lambda_L$ is +ve $D_{corr}\geq 3$, suggesting a low dimensional chaos.

$D_{corr}$ and $\lambda_L$ have been estimated for different data lengths to check the error on the estimations of the $D_{corr}$ and $\lambda_L$. The estimated $D_{corr}$ and $\lambda_L$ at different data lengths have been shown in Table~\ref{table:lyap}. Both and $\lambda_L$ and $D_{corr}$ tend to show stable results for higher data lengths (larger than 1000) as seen from Table~\ref{table:lyap}.

\begin{figure}[ht]
\includegraphics[width=8.5 cm]{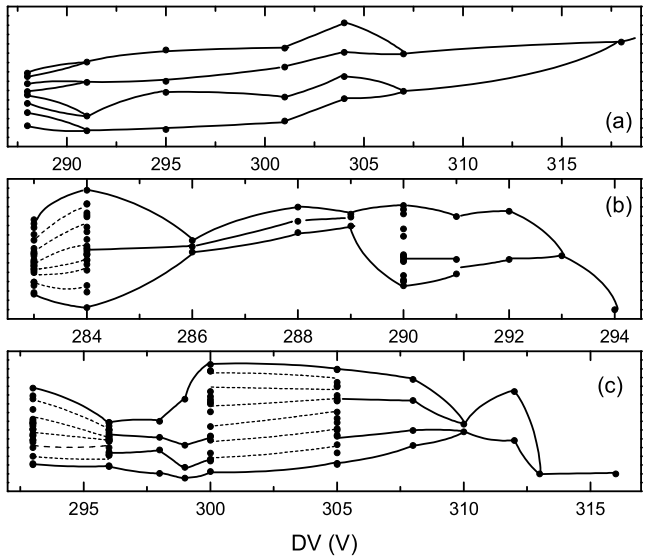}
\caption{ (a), (b), and (c) are the bifurcation diagram with DVs at pressures (a) 0.89, (b) 0.95, and (c) 1 mbar respectively.}
\label{fig:bifur}
\end{figure}

The evolution of the floating potential fluctuations have also been shown using the bifurcation diagram in Fig~\ref{fig:bifur}. At 0.89 mbar the system reaches a stable state through period subtraction [Fig~\ref{fig:bifur}(a)] with increasing DV. No chaos was observed in this case. For 0.95 mbar a chaotic region between DVs 283$-$284 V and an intermediate chaotic state have been observed at 290 V [Fig~\ref{fig:bifur}(b)]. We have two chaotic regions, one between 288$-$296 V and the other between 300$-$305 V at pressure 1 mbar [Fig~\ref{fig:bifur}(c)]. The figures show that inverse bifurcation probably takes place in the system and this has been observed for the first time in a glow discharge system.

\begin{figure}[ht]
\includegraphics[width=8.5 cm]{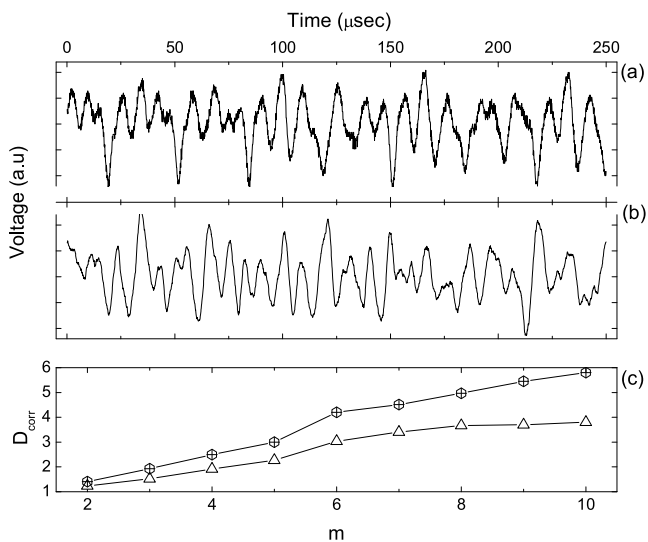}
\caption{Surrogate data using phase randomizing method: (a) original data; (b) surrogate data; (c) Correlation dimensions vs m for original ($\Delta$) and its surrogate data ($\oplus$).}
\label{fig:surgt}
\end{figure} 

In order to validate the above observations, we carried out the surrogate data analysis to detect the stochasticity or determinism in the system~\cite{physicaD:theiler,Bifur:Nakamura,ieee:small}. We have employed the correlation dimension as the test statistics for this method. The surrogate data has been generated by Phase Shuffled method, in which the phases of the signals are randomized by shuffling the fourier phases~\cite{chaos:dori,physicaD:theiler,Bifur:Nakamura}, and hence  the power spectrum (linear structure) is preserved, but the nonlinear structures are destroyed~\cite{Bifur:Nakamura}. As seen in Fig~\ref{fig:surgt}(c) The $D_{corr}$ for the original data saturates at higher m, whereas for the surrogate data  it increases with m as expected, since the surrogate data is supposed to be random in nature, and its $D_{corr}$ should be infinite~\cite{chaos:dori}. The saturation of $D_{corr}$ in our case indicates that the system dynamics is deterministic in nature.

\section{conclusions}
\label{section:discussion and conclusion}

Glow discharges are simple systems but exhibit exotic features depending on the configuration, initial parameters etc. Since various configurations are used in different applications like dusty plasma, plasma processing etc. it is necessary to understand the plasma dynamics of the these systems as much as possible. However the understanding of the complexities in the plasma dynamics is quite a challenging job as they arise from many degrees of freedom like different sources of free energy, different types of wave particle interaction and many other instabilities. In our present work nonlinear time series analysis has been used to quantify and differentiate complex and coherent processes at different parametric conditions. We also observed that the chaotic state driven by a driving force (DV) tends to be stabilized by some type of inverse bifurcation. This is also evident from the estimated $D_{corr}$ and $\lambda_L$ at different DVs. It is not fully understood as to why relaxation oscillations are being observed only at 0.89 mbar and not at 0.95  and 1.0 mar though there is not much difference in the three pressures. One probably has to carry out a detailed simulation to investigate the evolution of the fluctuations at different discharge voltages.

\section*{Acknowledgment}
We gratefully acknowledge the use of the software for calculating largest Lyapunov exponent by M.T. Rosenstein et al [http://www.physionet.org/physiotools/lyapunov/l1d2/]. We acknowledge the discussion of the results with Dr. S.K. Dana, IICB, India and Prof. P. Parmananda, Mexico.  We would also like to thank S.S.Sil, D. Das, and D. N. Debnath for their help during the experiment.

\end{document}